\documentclass[twocolumn,english]{revtex4-1}
\usepackage{lmodern}

\usepackage[T1]{fontenc}
\usepackage[latin9]{inputenc}
\usepackage{color}
\usepackage{amsmath}
\usepackage{graphicx}

\makeatletter

\providecommand{\tabularnewline}{\\}

\PassOptionsToPackage{caption=false}{subfig}

\@ifundefined{showcaptionsetup}{}{%
 \PassOptionsToPackage{caption=false}{subfig}}
\usepackage{subfig}
\makeatother

\usepackage{babel}
\begin{document}

\title{Two-site Bose-Hubbard model with nonlinear tunneling: classical and
quantum analysis}

\author{D. Rubeni, J. Links and P. S. Isaac}

\affiliation{School of Mathematics and Physics, The University of Queensland,
Brisbane, QLD 4072, Australia}

\author{A. Foerster}

\affiliation{Instituto de F\'isica da UFRGS, Av. Bento Gon\c{c}alves 9500, Agronomia,
Porto Alegre, RS, Brazil }
\begin{abstract}
The extended Bose-Hubbard model for a double-well potential with atom-pair
tunneling is studied. Starting with a classical analysis we determine
the existence of three different quantum phases: self-trapping, phase-locking
and Josephson states. From this analysis we built the parameter space
of quantum phase transitions between degenerate and non-degenerate
ground states driven by the atom-pair tunneling. Considering only
the repulsive case, we confirm the phase transition by the measure
of the energy gap between the ground state and the first excited state.
We study the structure of the solutions of the Bethe ansatz
equations for a small number of particles. An inspection 
of the roots for the ground state suggests a relationship  to
the physical properties of the system. By studying the energy gap
we find that the profile of the roots of the Bethe ansatz equations is 
related to a quantum phase transition.
\end{abstract}
\maketitle

\section{Introduction}

The Bose\textendash Hubbard model for a double-well potential has
been extensively studied since the experimental realization of Bose\textendash Einstein
condensates (BECs). This simple model can well describe the Josephson
oscillations and nonlinear self-trapping of BECs in a double-well
trap \cite{ALBIEZ} with weak atom\textendash atom interactions. Due
to its simplicity, this model has been investigated widely by many
authors using various methods, such as the Gross-Pitaevskii approximation
\cite{LEGGETT}, mean-field theory \cite{MILBURN,HINES}, the quantum
phase model \cite{ANGLIN} and the Bethe ansatz method \cite{ZHOU},
providing insights into many intriguing phenomena. For example, it
is well known that this model may present a \emph{Quantum Phase Transition}
(QPT) separating a delocalised from a self-trapped phase \cite{TONEL,PAN}.

However, strong interaction may fundamentally alter the tunnel configuration
and result in a correlated tunnelling, which was explored most recently
in the context of ultracold atoms \cite{FOLLING,ZOLLNER}. The tunnelling
dynamics of a few atoms loaded in a double-well trap has been studied
by varying the interaction strength from a weak to strong limit and
it was shown for the two-atom case that the tunnelling character changes
from Rabi oscillation to an atom-pair co-tunnelling process with increasing
interaction. A direct observation of the correlated tunnelling was
reported recently \cite{FOLLING} and theoretical analysis has also
been presented in terms of two-body quantum mechanics \cite{ZOLLNER}.
It was shown that the two-mode Bose-Hubbard model (TMBH) should be
modified by a nonlinear interaction-dependent tunnelling term in the
case of a large number of atoms \cite{ANANIKIAN}, which leads to
a considerable contribution to the tunnelling effect. In \cite{LIANG},
it was pointed out that the Bose\textendash Hubbard Hamiltonian, which
is valid in a relatively weak interaction regime, is not able to describe
the dynamics of atom-pair tunnelling and should be extended in the
strong interacting regime to include the atom\textendash atom interaction
of neighbouring lattice sites. In the model under consideration, a
novel atom-pair hopping term is included to describe the two-body
interaction recently reported experimental observation of correlated
tunnelling. There has been a great deal of effort devoted to this
subject recently \cite{LIU,LIU2,ZHU,DUTTA,WEN}.

In this paper, we adopt a Hamiltonian including the atom-pair tunnelling
term to describe BECs in a double well potential. The extended two-mode
Bose-Hubbard model (eTMBH) can be described by the following Hamiltonian
\begin{eqnarray}
H & = & U_{1}\hat{n}_{1}^{2}+U_{2}\hat{n}_{2}^{2}-\frac{1}{2}\Delta\left(\hat{n}_{1}-\hat{n}_{2}\right)-\frac{J}{2}\left(\hat{a}_{1}^{\dagger}\hat{a}_{2}+\hat{a}_{2}^{\dagger}\hat{a}_{1}\right)\nonumber \\
 &  & -\frac{\Omega}{2}\left(\hat{a}_{1}^{\dagger}\hat{a}_{1}^{\dagger}\hat{a}_{2}\hat{a}_{2}+\hat{a}_{2}^{\dagger}\hat{a}_{2}^{\dagger}\hat{a}_{1}\hat{a}_{1}\right),\label{Quantum Hamiltonian}
\end{eqnarray}
where $\left\{ \hat{a}_{j},\,\hat{a}_{j}^{\dagger}|\,j=1,\,2\right\} $
are the creation and annihillation operators for well $j$ associated,
respectively, with two bosonic Heisenberg algebras, and satisfying
the following commutation relations 

\[
\left[\hat{a}_{i},\,\hat{a}_{j}^{\dagger}\right]=\delta_{ij},\,\left[\hat{a}_{i},\,\hat{a}_{j}\right]=\left[\hat{a}_{i}^{\dagger},\,\hat{a}_{j}^{\dagger}\right]=0.
\]
Also $\hat{n}_{j}=\hat{a}_{j}^{\dagger}\hat{a}_{j}$ is the corresponding
boson number operator for each well. Since the Hamiltonian commutes with the total boson number
operator $\hat{n}=\hat{n}_{1}+\hat{n}_{2}$, the total number of bosons $n$
is conserved and it is convenient to restrict to a subspace
of constant $n$. The coupling $U_{j}$ provides the strength
of the scattering interaction between bosons in the well $j$ and
may be attractive $\left(U_{j}<0\right)$ or repulsive $\left(U_{j}>0\right)$.
The parameter $\Delta$ is the external potential which corresponds
to an asymmetry between the condensates, $J$ is the coupling for
the tunneling and $\Omega$ is a factor to describes the atom-pair
tunneling process. The change $J\rightarrow-J$ corresponds to the
unitary transformation $\hat{a}_{1}\rightarrow\hat{a}_{1}$, $\hat{a}_{2}\rightarrow-\hat{a}_{2}$,
while $\Delta\rightarrow-\Delta$ corresponds to $\hat{a}_{1}\leftrightarrow\hat{a}_{2}$.
Therefore we will restrict our analysis to the case of $J,\,\Delta\ge0$. 

Undertaking a classical analysis we obtain the fixed points
of the system in the large $n$ limit, and find three distinct
phases for the ground state. Under the right conditions
the system may undergo a QPT. The results for some particular cases
allow us to identify a parameter space of quantum phase transitions. We then
confirm that this parameter space is associated with quantum phase
transitions of the system through studies of the energy gap. 

Then we present the exact solution for this model using the Bethe
ansatz approach. By this method one can have access to the ground
state through the solution of a set of Bethe ansatz equations. A careful
observation of the behavior of solutions of these equations for the
ground state, as we vary some parameters of the Hamiltonian, suggests
a connection between the behavior of roots of the Bethe ansatz equations
and the physical behavior of such model. This is exactly what we expect
to happen in quantum phase transitions.

This paper is organized as follows: in the second section we analyze
the eTMBH model through bifurcations in a classical analysis. These are used to indicate
potential quantum phase transitions. We find the fixed points
for the special case $\Delta=0,\,U_{1}=U_{2}$ and build a parameter
space of phase transitions. A comparison is made between the classical
predictions and the energy gap. In the third section we present the
Bethe ansatz solution and investigate the distribution
of the roots of the Bethe ansatz equations for the ground state.
In the fourth section we summarize our results.

\section{Classical analysis}

We start our analysis with a semi-classical treatment. We study the
phase space of this system, in particular determining the fixed points.
It is found that for certain coupling parameters bifurcations of the
fixed points occur, and we can determine a parameter space diagram
which classifies the fixed points. 

For this second-quantized model, if the particle number $n$ is large
enough, the system can be well described in the classical approximation
\cite{WU}, where creation/annihillation operators can be replaced
by complex numbers $\left(n_{j},\,\theta_{j}\right)$ such as 
\[
\hat{a}_{j}\rightarrow e^{i\theta_{j}}\sqrt{n_{j}},\,\,\,\hat{a}_{j}^{\dagger}\rightarrow\sqrt{n_{j}}e^{-i\theta_{j}}.
\]
By introducing the canonically conjugate variables \emph{population
imbalance $z$} and \emph{phase difference} $\theta$, defined by
\[
z=\frac{1}{n}\left(n_{1}-n_{2}\right),\,\theta=\frac{n}{2}\left(\theta_{1}-\theta_{2}\right),
\]
the system can be described by the classical Hamiltonian

\begin{eqnarray}
\text{\textit{\ensuremath{\mathcal{H}}}} & = & \frac{nJ}{4}\left(\lambda\left(1+z^{2}\right)-\gamma\left(1-z^{2}\right)\cos\left(4\theta/n\right)\right.\nonumber \\
 &  & \left.-2\sqrt{1-z^{2}}\cos\left(2\theta/n\right)-2\beta z\right),\label{Classical Hamiltonian}
\end{eqnarray}
where
\[
\lambda=\frac{n}{J}\left(U_{1}+U_{2}\right),\,\beta=\frac{n}{J}\left(\frac{\Delta}{n}-U_{1}+U_{2}\right)\,\textrm{and}\,\gamma=\frac{n\Omega}{J}
\]
are the coupling parameters. Hamilton\textquoteright s equations of motion are given by
\begin{eqnarray}
\dot{z} & = & -J\sin\left(2\theta/n\right)\left(2\gamma\cos\left(2\theta/n\right)-2\gamma z^{2}\cos\left(2\theta/n\right)\right.\nonumber \\
 &  & \left.+\sqrt{1-z^{2}}\right)\label{z derivative}
\end{eqnarray}
\begin{equation}
\dot{\theta}=\frac{nJ}{2}\left(-\beta+\gamma z\cos\left(4\theta/n\right)+\frac{z\cos\left(2\theta/n\right)}{\sqrt{1-z{}^{2}}}+\lambda z\right)\label{theta derivative}
\end{equation}
In the limit $\gamma\rightarrow0$
we recover the equations of motion of the TMBH \cite{RAGHAVAN}. The
fixed points can be readily derived from the condition $\dot{z}=\dot{\theta}=0$.
Due to periodicity of the solutions, below we restrict to $2\theta/n\in\left[-\pi,\,+\pi\right]$.
This leads to the following classification:
\begin{itemize}
\item $\theta=0$ and $z$ is a solution of 
\begin{equation}
-\beta+z\left(\gamma+\lambda\right)=-\frac{z}{\sqrt{1-z{}^{2}}},\label{condition1}
\end{equation}
which has one solution for $\lambda+\gamma\geq-1$ while may have
one, two or three solutions for $\lambda+\gamma<-1$ . In Figure \ref{fig: Graphical Solutions}
we present a graphical solution of (\ref{condition1}). 
\item $2\theta/n=\pm\pi$ and $z$ is a solution of 
\begin{equation}
-\beta+z\left(\gamma+\lambda\right)=\frac{z}{\sqrt{1-z{}^{2}}}.\label{condition2}
\end{equation}
This equation has one solution for $\lambda+\gamma\leq1$ and has
either one, two or three real solutions for $\lambda+\gamma>1$.
\item $z=\beta/\left(\lambda-\gamma\right)$ and $\theta$ is a solution
of
\begin{equation}
\cos\left(2\theta/n\right)=\frac{-1}{2\gamma\sqrt{1-\left(\frac{\beta}{\lambda-\gamma}\right)^{2}}},\label{condition3}
\end{equation}
which has two real solutions for $\gamma\notin\left[-1/2,\,1/2\right]$
and $\left|\lambda-\gamma\right|\geq2|\beta\gamma|\left(4\gamma^{2}-1\right){}^{-1/2}$.
\end{itemize}
\begin{center}
\begin{figure}
\begin{centering}
\includegraphics{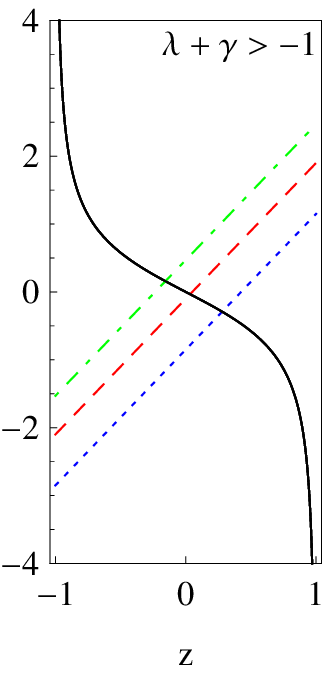}~~~\includegraphics{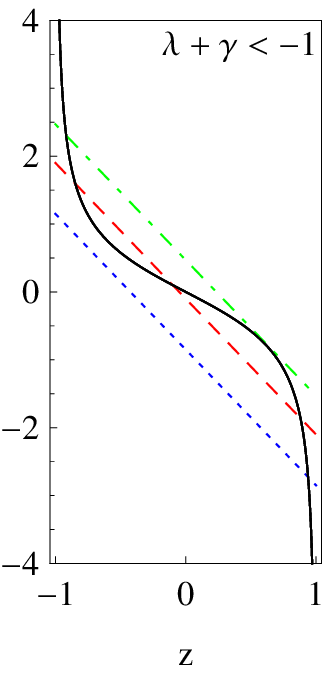}
\par\end{centering}

\caption{\label{fig: Graphical Solutions}Graphical solution of equation (\ref{condition1}).
The crossing between the straight line (left hand side of eq. (\ref{condition1}))
and the curve (right hand side of eq. (\ref{condition1})) for different
values of $\lambda+\gamma$ and $\beta$ represents the solution(s)
for each case. There is just one solution on the left $\left(\lambda+\gamma\geq-1\right)$
while there are either one, two or three solutions on the right $\left(\lambda+\gamma<-1\right)$.}
\end{figure}

\par\end{center}

From the equations (\ref{condition1}) and (\ref{condition2}) we
can determine that there are fixed point bifurcations for certain
choices of the coupling parameters. These bifurcations allow us to
divide the coupling parameter space in three regions. A standard analysis
shows the boundary between the regions obey the relation 
\begin{equation}
\lambda+\gamma=\pm\left(1+\left|\beta\right|^{\frac{2}{3}}\right)^{\frac{3}{2}}\label{boundaries 1}
\end{equation}
(see \cite{LINKS} for details). Eq. (\ref{boundaries 1}) leads to a partition of the parameter space into three regions, depicted
in Figure \ref{fig:ParameterSpaceA1}. In the absence
of the external potential, i.e. $\beta=0$, we have a fixed
point bifurcation given by $\lambda=\pm1-\gamma$. See Figure \ref{fig:ParameterSpaceA2}. Irrespective of the nature of the bifurcation, 
it has been observed in the classical analysis
\cite{HINES2,SCHNEIDER} that fixed points 
can be used to identify quantum phase transitions. This model therefore becomes a promising candidate
to study.

The conditions for existence of solutions to equation (\ref{condition3})
allow us to build a parameter space diagram as depicted in Figure
\ref{fig:ParameterSpaceB1}. The boundary between regions satisfies
the relation
\begin{equation}
\lambda-\gamma=\pm2|\beta\gamma|\left(4\gamma^{2}-1\right){}^{-1/2}\label{boundaries 2}
\end{equation}

\begin{center}
\begin{figure}
\subfloat[\label{fig:ParameterSpaceA1}]{\includegraphics{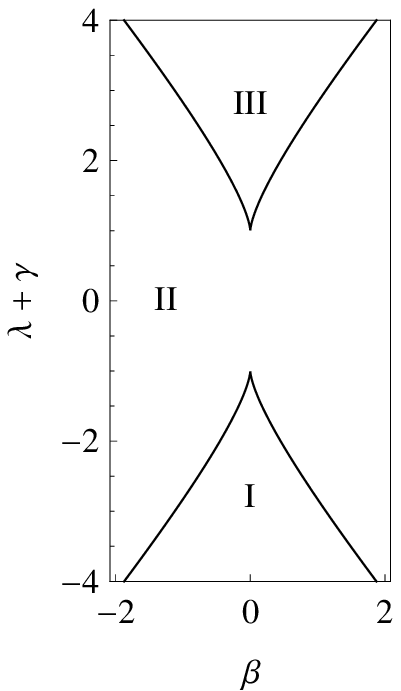}

}~\subfloat[\label{fig:ParameterSpaceA2}]{\begin{centering}
\includegraphics{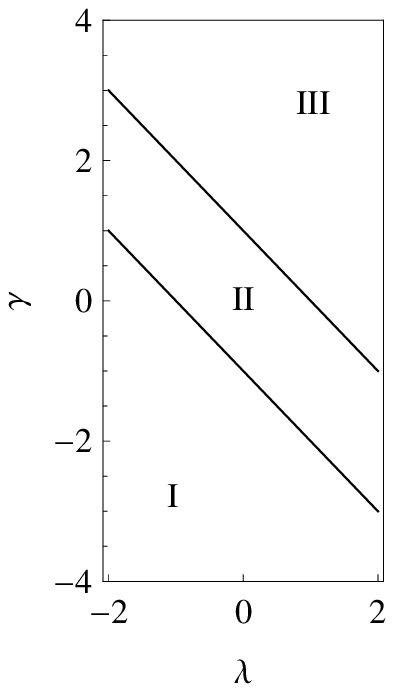}
\par\end{centering}

}\\
\subfloat[\label{fig:ParameterSpaceB1}]{\includegraphics{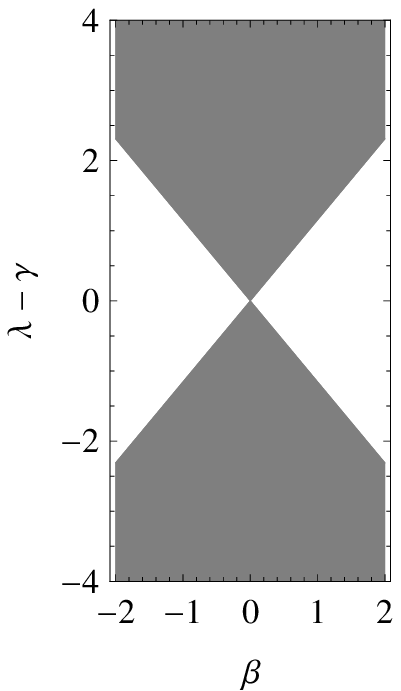}}~\subfloat[\label{fig:ParameterSpaceB2}]{\centering{}\includegraphics{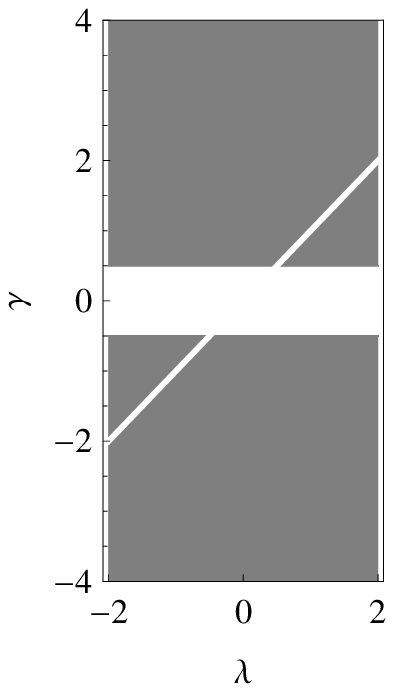}}

\caption{Coupling parameter space diagrams characterizing the solutions for the fixed points 
$\dot{z}=\dot{\theta}=0$.
(a) Parameter space for equations (\ref{condition1}) and (\ref{condition2})
with $\beta\protect\neq0$. The boundaries between the regions are
given by equations (\ref{boundaries 1}). At the boundary between
the regions I and II there are two solutions for $\theta=0$ and one
solution for $2\theta/n=\pm\pi$, while there is one solution for
$\theta=0$ and two solutions for $2\theta/n=\pm\pi$ at the boundary
between regions II and III. (b) Parameter space for equations (\ref{condition1})
and (\ref{condition2}) with $\beta=0$. The boundaries between the
regions obey the equations $\gamma=\pm1-\lambda.$ In both cases,
there are three solutions for $\theta=0$ and one solution $2\theta/n=\pm\pi$
in the region I; In the region II we have one solution for $\theta=0$
and one solution for $2\theta/n=\pm\pi$; In the region III there
is one solution for $\theta=0$ and three solutions for $2\theta/n=\pm\pi$.
(c) Example of parameter space for equation (\ref{condition3}) with
$\beta\protect\neq0$. This equation only has one solution for the
values of parameters that lie within the shaded area, with boundaries
given by (\ref{boundaries 2}), and $\left|\gamma\right|>1/2$. (d)
Parameter space for equation (\ref{condition3}) with $\beta=0$.
This equation only has a solution for the values of parameters that
lie within the light gray area, with $\left|\gamma\right|>\frac{1}{2}$
and $\gamma\protect\neq\lambda$. }
\end{figure}

\par\end{center}

\subsection{Fixed points and eigenstates for $\beta=0$}

In the following we will study the solutions of the fixed point equations
(\ref{condition1}), (\ref{condition2}) and (\ref{condition3}) with
$\beta=0$ by the consideration of two main reasons: (i) nonzero values
of $\Delta$ do not significantly alter the behavior of the system,
just shifting the energy levels \cite{TONEL} and (ii) much of the
experimental realizations with these systems are made on the condition
of zero external potential and equal interaction between atoms in
each well \cite{FOLLING}. In Figure \ref{fig:ParameterSpaceA2} we
see the parameter space diagram for equations (\ref{condition1})
and (\ref{condition2}) with $\beta=0$, while Figure \ref{fig:ParameterSpaceB2}
shows the parameter space diagram for equation (\ref{condition3})
for $\beta=0$. 

It has been demonstrated that the fixed points of phase-space level
curves are the points of extreme energy corresponding to eigenstates
of the system \cite{JIE}. Since the fixed point bifurcations change
the topology of the level curves, qualitative differences can be observed
between each of the three regions. For further analysis, it is useful
to assign to each fixed point $\left(\theta_{FP},\,z_{FP}\right)$
a point $P_{j}$ in the phase space as follows:

\[
\begin{cases}
P_{1}\rightarrow\left(0,\,0\right)\\
P_{2}\rightarrow\left(0,\,\pm\sqrt{1-1/\left(\lambda+\gamma\right)^{2}}\right)\\
P_{3}\rightarrow\left(\pm\textrm{arcsec}\left(-2\gamma\right),\,0\right)\\
P_{4}\rightarrow\left(\pm\pi,\,0\right)\\
P_{5}\rightarrow\left(\pm\pi,\,\pm\sqrt{1-1/\left(\lambda+\gamma\right)^{2}}\right)
\end{cases}
\]

\begin{figure}
\begin{centering}
\includegraphics{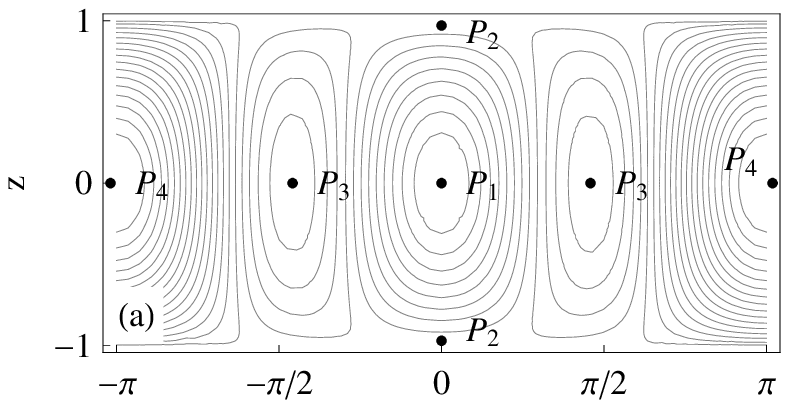}\\
\includegraphics{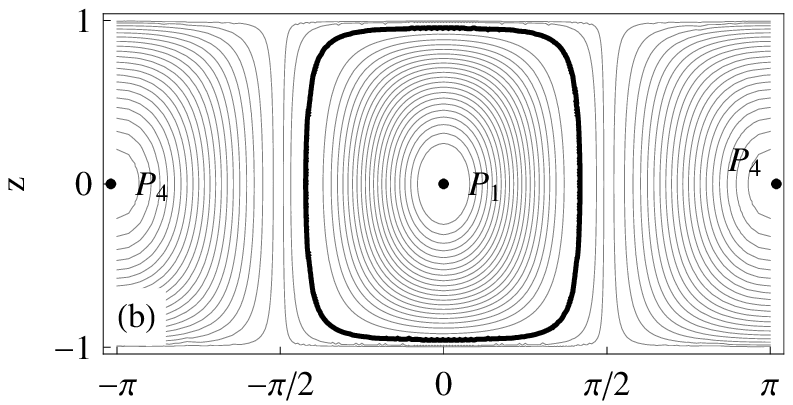}\\
\includegraphics{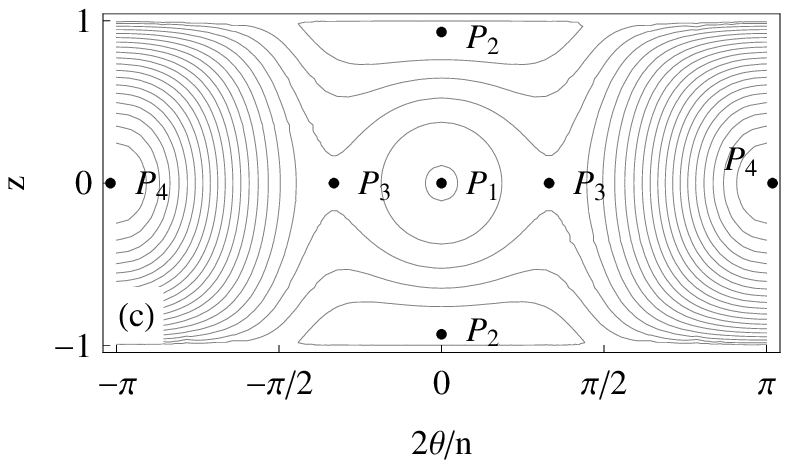}
\par\end{centering}

\caption{\label{fig: LevelCurvesRI}Level curves of the classical Hamiltonian
in region I. The points $\left\{ P_{1},\,...,\,P_{4}\right\} $ denote
the fixed points of the Hamiltonian. (a) the parameter values are
$n=100$, $\lambda=-2$, $J=1$ and $\gamma=-4$. There is a local
maximum at $P_{1}$ and saddle points at $P_{2}$. Global minima are
at $P_{3}$, while $P_{4}$ are global maxima. In (b) the parameter
values are $n=100$, $\lambda=-2$, $J=1$ and $\gamma=-2$. A ``ring''
emerges as the global minimum. The local and global maxima still occur
at $P_{1}$ and $P_{4}$, respectively. (c) Now the parameter values
are $n=100$, $\lambda=-2$, $J=1$ and $\gamma=-1$. Global minima
are at $P_{2}$. There are saddle points at $P_{3}$, a local maximum
point at $P_{1}$ and global maxima at $P_{4}$.}
\end{figure}

Figure \ref{fig: LevelCurvesRI} shows the typical character of the
level curves in region I. There are three fixed points for $\theta=0$
and one fixed point for $2\theta/n=\pm\pi$. When $\gamma<\lambda$
the ground state is associated with the fixed points $P_{3}$. These
two states are called \emph{phase-locking states} with zero population
imbalance and tunable relative phase unequal to $0$ or $\pi$ - see
Figure 3a. This phase-locking state was also identified in \cite{LIANG}.
Highest energetic states corresponds to the fixed points $P_{4}$.
At $\gamma=\lambda$ the system changes to a special state: the ground
state is over a ``ring'' instead a of point, as depicted in Figure
3b. This is a transition state, since any small changes in the values
of $\lambda$ and $\gamma$ alter its nature. When $\gamma>\lambda$
there are an abrupt change in the ground state: the minima energy
levels moves towards the fixed points $P_{2}$. We denote \emph{self-trapping
states} as those eigenstates whose corresponding fixed points have
a nonzero population imbalance, $z\neq0$, as depicted in Figure 3c.
Therefore, now the ground state is a degenerate self-trapping state.
This means that at $\gamma=\lambda$ the system undergoes a QPT from
degenerate phase-locking states to degenerate self-trapping states.
Further changes in the coupling parameters modify the fixed point
configuration, but no longer alter the nature of the ground state.
Table \ref{tab: I} provides a detailed classification for all the
fixed points in region I as the parameters $\lambda$ and $\gamma$
change.
\begin{ruledtabular}
\begin{center}
\begin{table}
\begin{centering}
\begin{tabular}{cccccc}
Region I & ~~~~$P_{1}$~~~ & ~~~~$P_{2}$~~~ & ~~~~$P_{3}$~~~ & ~~~~$P_{4}$~~~ & ~~~~$P_{5}$~~~\tabularnewline
\hline 
$\gamma<\lambda$ & \emph{lmax} & \emph{sp} & \emph{GS} & \emph{HES} & \emph{-----}\tabularnewline
$-1/2>\gamma>\lambda$ & \emph{lmax} & \emph{GS} & \emph{sp} & \emph{HES} & \emph{-----}\tabularnewline
$-1/2<\gamma<1/2$ & \emph{sp} & \emph{GS} & \emph{-----} & \emph{HES} & \emph{-----}\tabularnewline
$\gamma>1/2$ & \emph{sp} & \emph{GS} & \emph{HES} & \emph{sp} & -----\tabularnewline
\end{tabular}
\par\end{centering}

\caption{\label{tab: I}Configuration of fixed points and associated states
in region I. In this table, GS means \emph{Ground State}, lmax is
a\emph{ local maxima}, sp is a \emph{saddle point} while HES refers
to the \emph{Highest Excited State.} }
\end{table}

\par\end{center}
\end{ruledtabular}

Figure \ref{fig: LevelCurvesRII}a illustrates the configuration of
the fixed points when the coupling parameters are tuned to cross over
from region I into region II. There is one fixed point for $\theta=0$
and one for $2\theta/n=\pm\pi$. If $\gamma>-1/2$ the fixed point
$P_{1}$ becomes associated with the ground state, with zero population
imbalance and zero relative phase, with the presence of tunnelling
of atoms between the wells because of the weak interaction. We call
this state a \emph{Josephson state}. Therefore, when crossing the
boundary $\gamma=-1-\lambda$, the system undergoes a QPT to a non-degenerate
Josephson state. Highest excited states are related to the global
maxima at $P_{3}$. If $\gamma<-1/2$, there is another QPT: the global
minima, related to degenerate phase-locking states, emerges at $P_{3}$
- see Figure 4b. Highest energy states are associated with the fixed
point $P_{4}$ for any $\lambda<1/2$. Table \ref{tab: II} summarizes
how the fixed point configurations change along with $\lambda$ and
$\gamma$.

\begin{center}
\begin{figure}
\begin{centering}
\includegraphics{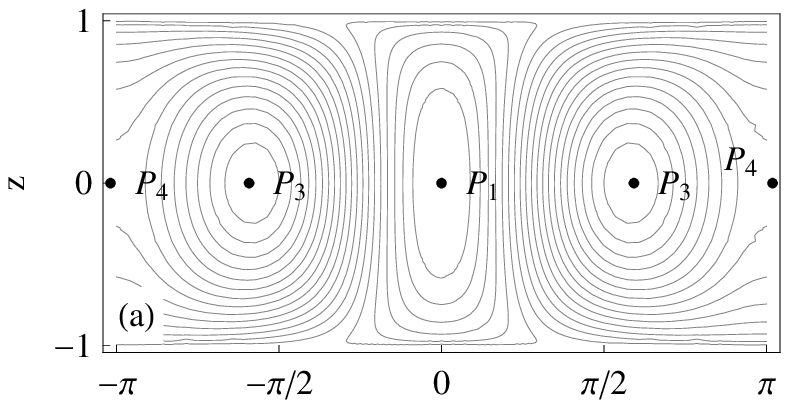}\\
\includegraphics{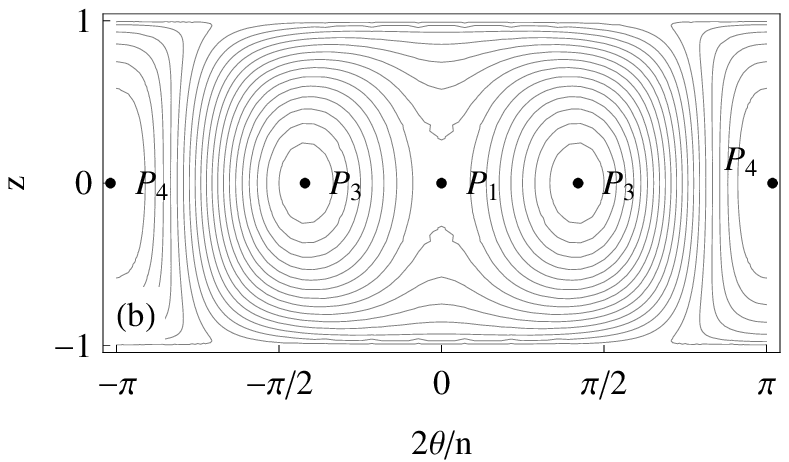}
\par\end{centering}

\caption{\label{fig: LevelCurvesRII}Typical level curves of the classical
Hamiltonian in region II. The points $\left\{ P_{1},\,...,\,P_{4}\right\} $
denote the fixed points of the Hamiltonian. (a) The parameter values
are $n=100$, $\lambda=-2$, $J=1$ and $\gamma=2$. There is a global
minimum at $P_{1}$, global maxima is at $P_{3}$, while $P_{4}$
are saddle points. In (b) the parameter values are $n=100$, $\lambda=2$,
$J=1$ and $\gamma=-2$. Now the fixed point $P_{1}$ turns into a
saddle point, while there are global minima at $P_{3}$ and global
maxima emerge at $P_{4}$. }
\end{figure}

\par\end{center}
\begin{ruledtabular}
\begin{center}
\begin{table}
\begin{centering}
\begin{tabular}{cccccc}
Region II & ~~~~$P_{1}$~~~ & ~~~~$P_{2}$~~~ & ~~~~$P_{3}$~~~ & ~~~~$P_{4}$~~~ & ~~~~$P_{5}$~~~\tabularnewline
\hline 
$-1/2>\gamma$ & \emph{sp} & ----- & \emph{GS} & \emph{HES} & -----\tabularnewline
$-1/2<\gamma<1/2$ & \emph{GS} & ----- & ----- & \emph{HES} & -----\tabularnewline
$\gamma>1/2$ & \emph{GS} & ----- & \emph{HES} & \emph{sp} & ---\tabularnewline
\end{tabular}
\par\end{centering}

\caption{\label{tab: II}Configuration of fixed points and associated states
in region II. In this table, GS means \emph{Ground State}, sp is a
\emph{saddle point} while HES refers to the \emph{Highest Excited
State.} }
\end{table}

\par\end{center}
\end{ruledtabular}

On crossing the parameter space boundary to region III, the fixed
point configuration change again: there is one fixed point for $\theta=0$
and three fixed points for $2\theta/n=\pm\pi$. The ground state of
the system may be associated with $P_{3}$ as a degenerate phase-locking
state if $\gamma<-1/2$. New fixed points emerge at $P_{5}$ as highest
energetic states. If $\gamma>-1/2$, the global minima changes to
$P_{1}$ and becomes associated with a non-degenerate Josephson state.
Therefore the line $\gamma=-1/2$ defines the boundary for a QPT -
see Figure \ref{fig: LevelCurvesRIII}a and Figure \ref{fig: LevelCurvesRIII}b. 

\begin{center}
\begin{figure}
\begin{centering}
\includegraphics{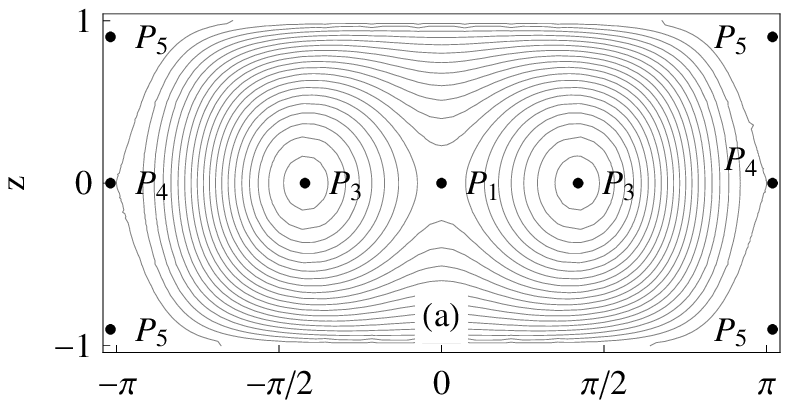}\\
\includegraphics{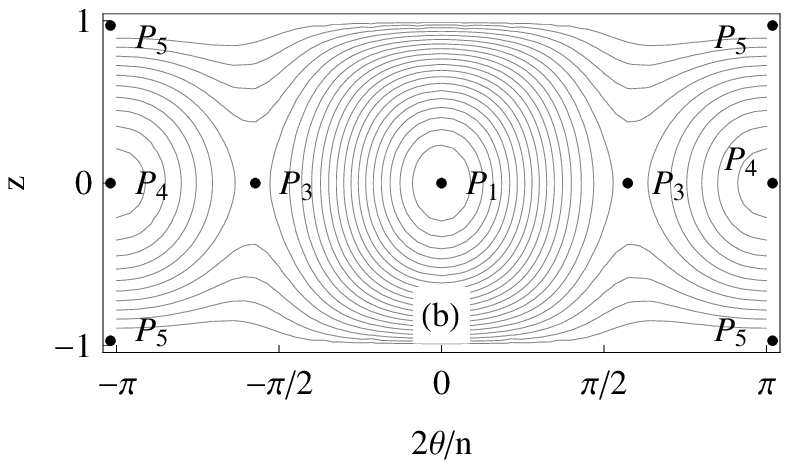}
\par\end{centering}

\caption{\label{fig: LevelCurvesRIII}Typical level curves of the classical
Hamiltonian in region III. The points $\left\{ P_{1},\,...,\,P_{5}\right\} $
denote the fixed points of the Hamiltonian. In (a) the parameter values
are $n=100$, $\lambda=4$, $J=1$ and $\gamma=-2$. In this scenario
$P_{1}$ is a saddle point and the global minima are at $P_{3}$.
Highest energy levels appears at $P_{5}$. In (b) the parameter values
are $n=100$, $\lambda=4$, $J=1$ and $\gamma=2$. Now the global
minima move towards $P_{1}$ while $P_{3}$ become saddle points.
The fixed points $P_{4}$ are local minima and the global maxima still
at $P_{5}$.}
\end{figure}

\par\end{center}
\begin{ruledtabular}
\begin{center}
\begin{table}
\begin{centering}
\begin{tabular}{cccccc}
Region III & ~~~~$P_{1}$~~~ & ~~~~$P_{2}$~~~ & ~~~~$P_{3}$~~~ & ~~~~$P_{4}$~~~ & ~~~~$P_{5}$~~~\tabularnewline
\hline 
$-1/2>\gamma$ & \emph{sp} & ----- & \emph{GS} & \emph{sp} & \emph{HES}\tabularnewline
$-1/2<\gamma<1/2$ & \emph{GS} & ----- & ----- & \emph{sp} & \emph{HES}\tabularnewline
$1/2<\gamma<\lambda$ & \emph{GS} & ----- & \emph{sp} & \emph{lmin} & \emph{HES}\tabularnewline
$\gamma>\lambda$ & \emph{GS} & ----- & \emph{HES} & \emph{lmin} & \emph{sp}\tabularnewline
\end{tabular}
\par\end{centering}

\caption{\label{tab:Tab. 3}Configuration of fixed points and associated states
in region III. In this table, GS means \emph{Ground State}, lmin is
a\emph{ local m}inima, sp is a \emph{saddle point} while HES refers
to the \emph{Highest Excited State.} }
\end{table}

\par\end{center}
\end{ruledtabular}

The above discussion gives a general qualitative description of the
behaviour of the classical system in terms of the three regions identified
in the parameter space. Properties of eigenstates as highlighted in Tables \ref{tab: I}, \ref{tab: II}, and \ref{tab:Tab. 3}
enables us to depict the quantum phase transition diagram shown in Figure
\ref{fig:ParameteSpaceQPT}. The parameter space $\left(\lambda,\,\gamma\right)$
is divided into three regions: self-trapping, Josephson, and phase-locking
phases. 

\begin{center}
\begin{figure}
\begin{centering}
\includegraphics{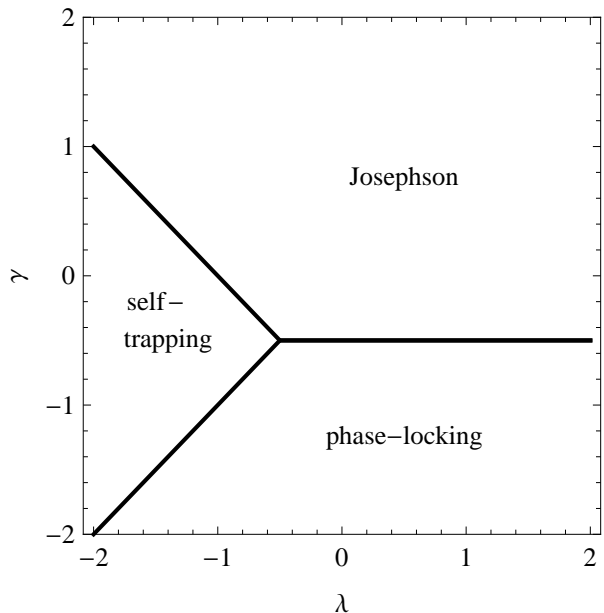}
\par\end{centering}

\caption{\label{fig:ParameteSpaceQPT}Parameter space for quantum phase transitions.
The boundary between Josephson and phase-locking states is given by
$\gamma=-1/2$. The system undergoes a QPT from phase-locking states
to self-trapping states by crossing the boundary $\gamma=\lambda$,
while the limit between the Josephson phase and the self-trapping
phase is determined by the line $\gamma=-1-\lambda$. The threshold
coupling occur at $\left(\gamma,\,\lambda\right)=\left(-1/2,\,-1/2\right)$.}
\end{figure}

\par\end{center}

In the next section we restrict ourselves to study the case $\lambda>0$
and check the presence of a phase transition as predicted by the phase
transition diagram studying the behaviour of the energy gap.

\subsection{Energy gap}

Consider the energy gap between the first excited state (FES) and
the ground state (GS),

\begin{equation}
\Delta E=E_{FES}-E_{GS}.\label{Gap}
\end{equation}
The values of the parameters for which the gap goes to zero identifies
the location of the QPT \cite{SACHDEV}. Using numerical diagonalization
of the Hamiltonian (\ref{Quantum Hamiltonian}), in Fig. \ref{fig:EnergyGapRepulsivea}
we plot the energy gap as a function of the coupling $\gamma$, for
$\lambda>0$ and different values of $n$. We observe that as $n$
increases the energy gap decreases and the coupling approaches the
point $\gamma=-1/2$. Fig. \ref{fig:EnergyGapRepulsiveb} shows similar
results for fixed $n$ and varying $\lambda$. We observe that the
occurrence of the vanishing of the gap, determining the QPT, fits
well with the predicted boundary separating Josephson and phase-locking
regions given by $\gamma=-1/2$.

\begin{center}
\begin{figure}
\begin{centering}
\subfloat[\label{fig:EnergyGapRepulsivea}]{\begin{centering}
\includegraphics{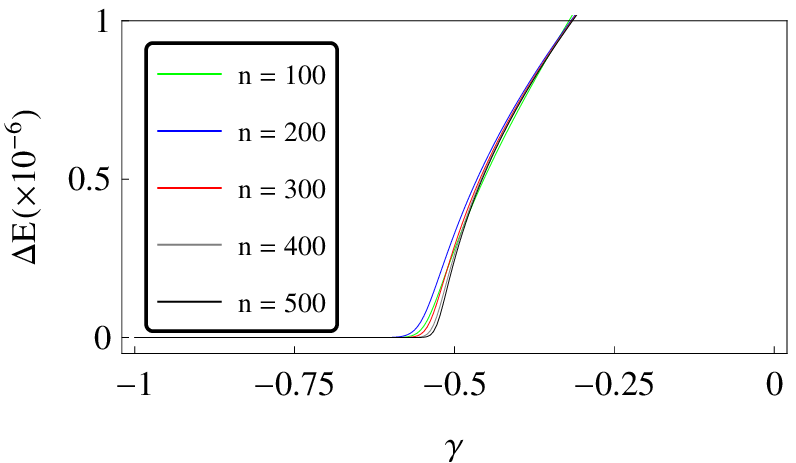}
\par\end{centering}

}\\
\subfloat[\label{fig:EnergyGapRepulsiveb}]{\centering{}\includegraphics{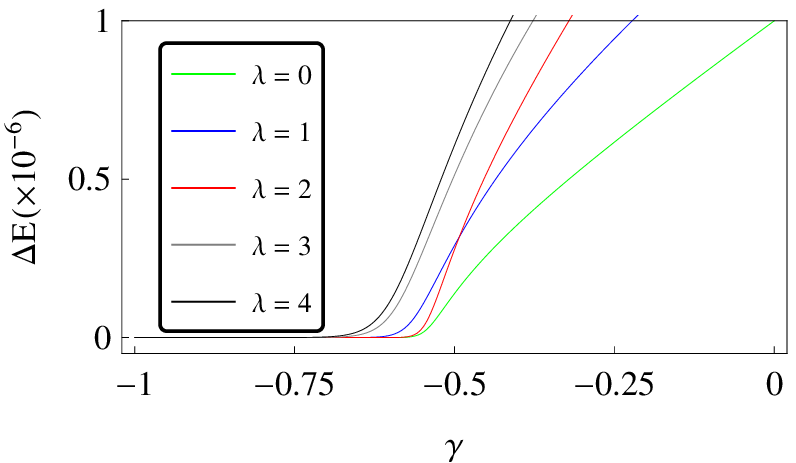}}
\par\end{centering}

\caption{Energy gap between the first excited state and the ground state as
a function of $\gamma=n\Omega/J$ for (a) different values of $n$
and $\lambda=2$ and (b) for different values of $\lambda$ and $n=100$.
The values of the parameters are $J=1$ and $\beta=0$. These results
indicate that the points at which the gap closes lie approximately
on the line $\gamma=-1/2$. }
\end{figure}

\par\end{center}

\section{Bethe ansatz solution}

To obtain the exact solution of the eTMBH model, we follow the work of Enol\textquoteright skii, Kuznetsov and
Salerno \cite{ENOLSKII}. Starting with the Jordan-Schwinger realisation
of the $su(2)$ algebra: 

\[
\hat{S}^{+}\rightarrow\hat{a}_{1}^{\dagger}\hat{a}_{2},\,\,\hat{S}^{-}\rightarrow\hat{a}_{2}^{\dagger}\hat{a}_{1},\,\,\hat{S}^{z}\rightarrow\frac{\hat{n}_{1}-\hat{n}_{2}}{2}
\]
we may write the Hamiltonian (\ref{Quantum Hamiltonian}) as

\begin{eqnarray}
H& = & \frac{k}{8}\hat{n}^{2}+\frac{k}{2}\left(\hat{S}^{z}\right)^{2}+\alpha\hat{S}^{z}-\frac{1}{2}J\left(\hat{S}^{+}+\hat{S}^{-}\right)\nonumber \\
 &  & -\frac{1}{2}\Omega\left[\left(\hat{S}^{+}\right)^{2}+\left(\hat{S}^{-}\right)^{2}\right]\label{Transformed Hamiltonian}
\end{eqnarray}
with $\hat{n}=\hat{n}_{1}+\hat{n}_{2},\,\,k=2\left(U_{1}+U_{2}\right)$
and $\alpha=\left(U_{1}-U_{2}\right)n-\Delta$. Note that
\begin{equation}
\lambda=\frac{kn}{2J},\,\beta=-\frac{\alpha}{J}.\label{parameter correspondence}
\end{equation}
If we consider the differential realization of $\textrm{su}\left(2\right)$
operators,
\[
\hat{S}^{+}\rightarrow nu-u^{2}\frac{d}{du},\,\,\hat{S}^{-}\rightarrow\frac{d}{du},\,\,\hat{S}^{z}\rightarrow u\frac{d}{du}-\frac{n}{2}
\]
the Hamiltonian (\ref{Transformed Hamiltonian}) can be written as
\begin{equation}
H=A\left(u\right)\frac{d^{2}}{du^{2}}+B\left(u\right)\frac{d}{du}+C\left(u\right)\label{Differential Hamiltonian}
\end{equation}
with

\begin{eqnarray*}
A\left(u\right) & = & \frac{k}{2}u^{2}-\frac{\Omega}{2}\left(u^{4}+1\right)\\
B\left(u\right) & = & \frac{1}{2}\left\{ J\left(u^{2}-1\right)+\left[k\left(1-n\right)+2\alpha\right]u\right.\\
 &  & \left.-2\Omega\left(1-n\right)u^{3}\right\} \\
C\left(u\right) & = & \frac{k}{4}n^{2}-\frac{\alpha}{2}n-\frac{J}{2}nu-\frac{\Omega}{2}n\left(n-1\right)u^{2}
\end{eqnarray*}
Solving for the spectrum of the Hamiltonian is then equivalent to
solving the eigenvalue equation 
\begin{equation}
HQ(u)=EQ(u)\label{eigenvalue equation}
\end{equation}
where $H$ is represented by (\ref{Differential Hamiltonian}) and
$Q(u)$ is a polynomial function of $u$ of order $n$. Next, 
express $Q(u)$ in terms of its roots ${\upsilon_{j}}$:
\[
Q\left(u\right)=\prod_{j=1}^{n}\left(u-\upsilon_{j}\right)
\]
Evaluating (\ref{eigenvalue equation}) at $u=\upsilon_{l}$ for each
$l$ leads to the set of Bethe ansatz equations (BAE)
\begin{eqnarray}
 &  & \frac{-J\left(\upsilon_{l}^{2}-1\right)+\left(k\left(1-n\right)+2\alpha\right)\upsilon_{l}-2\Omega\left(1-n\right)\upsilon_{l}^{3}}{k\upsilon_{l}^{2}-\Omega\left(\upsilon_{l}^{4}+1\right)}\nonumber \\
 &  &\qquad  =\sum_{j\neq l}^{n}\frac{2}{\upsilon_{j}-\upsilon_{l}}\label{BAE}
\end{eqnarray}

\begin{center}
\begin{figure}
\centering{}\subfloat[]{\begin{centering}
\includegraphics{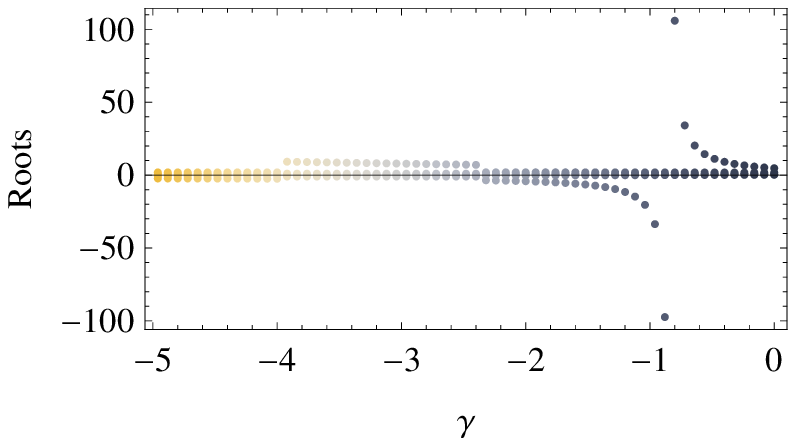}
\par\end{centering}

}\\
\subfloat[]{\centering{}\includegraphics{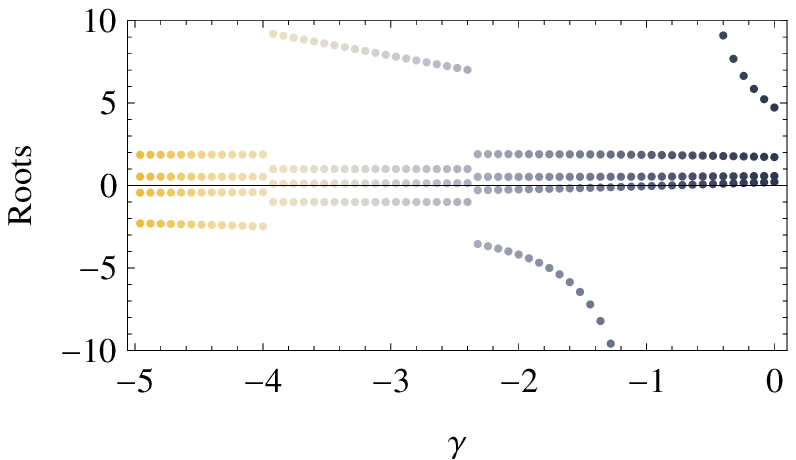}}\\
\subfloat[]{\centering{}\includegraphics{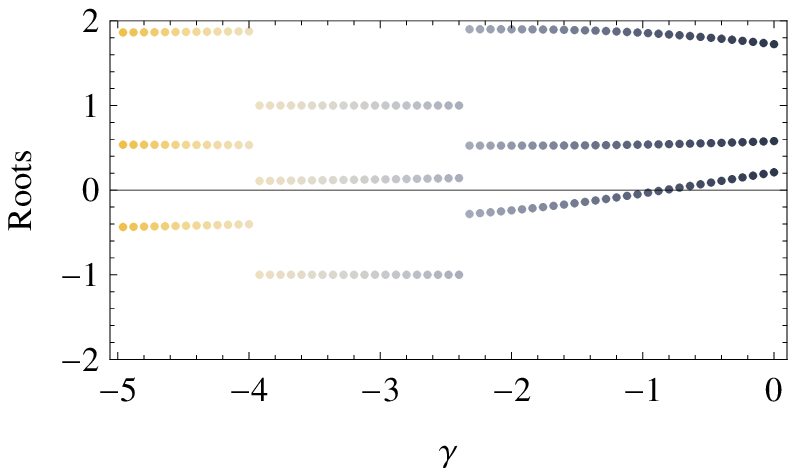}}\caption{\label{fig:BAEsolutionN4}Solutions of BAE (\ref{BAE}) for the
ground state considering the particular case $n=4$, $k=1$ and $J=1$
and different values of $\gamma$. The set of points with the same
color is the solution of the BAE for a given value of $\gamma$. In
(a), (b) and (c) we look at the same set of solutions in different
scales. There are a abrupt change in the roots distribution occurring
at $\gamma\simeq-2.38$ and $\gamma\simeq-4.01$.}
\end{figure}

\par\end{center}
Writing the asymptotic expansion

\[
Q\left(u\right)\sim u^{n}-u^{n-1}\sum_{j=1}^{N}\upsilon_{j}+u^{n-2}\sum_{j=1}^{n-1}\sum_{l=j+1}^{n}\upsilon_{j}\upsilon_{l}
\]

\begin{center}
\begin{figure*}[t]
\begin{centering}
\subfloat[]{\begin{centering}
\includegraphics{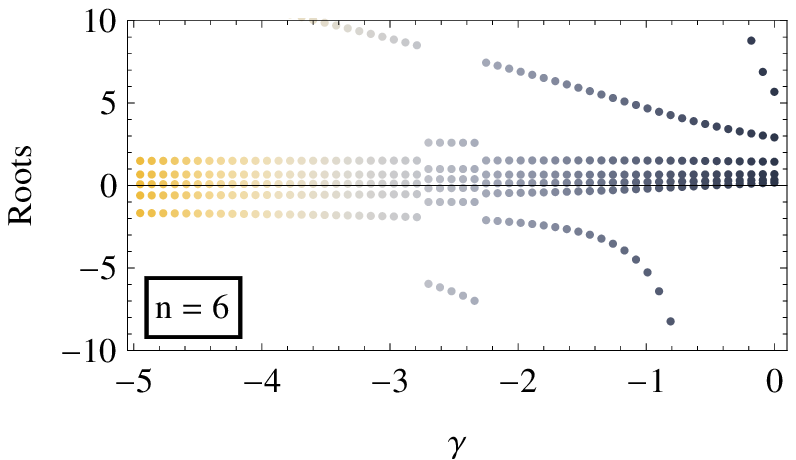}
\par\end{centering}

}~~~~~\subfloat[]{\centering{}\includegraphics{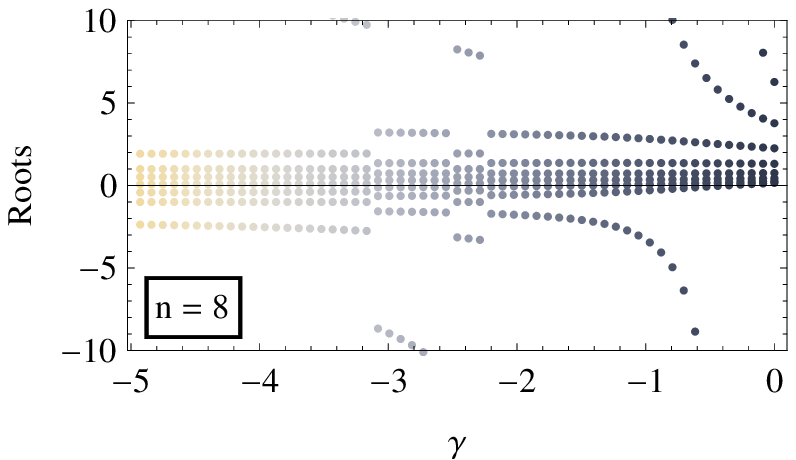}}\\
\subfloat[]{\centering{}\includegraphics{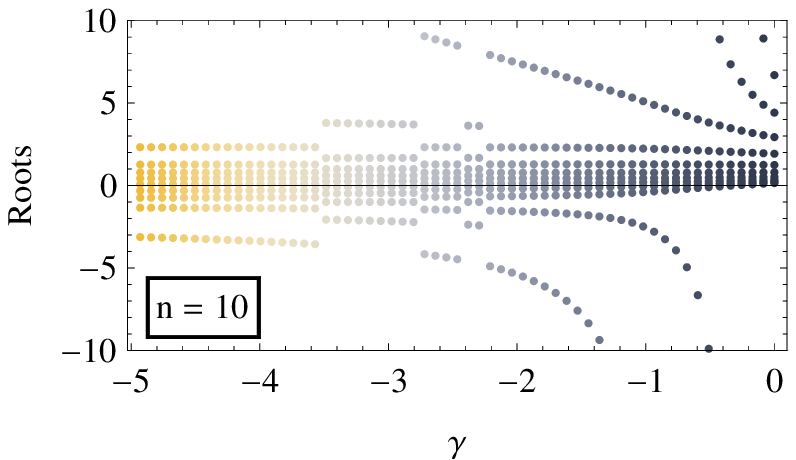}}~~~~~\subfloat[]{\centering{}\includegraphics{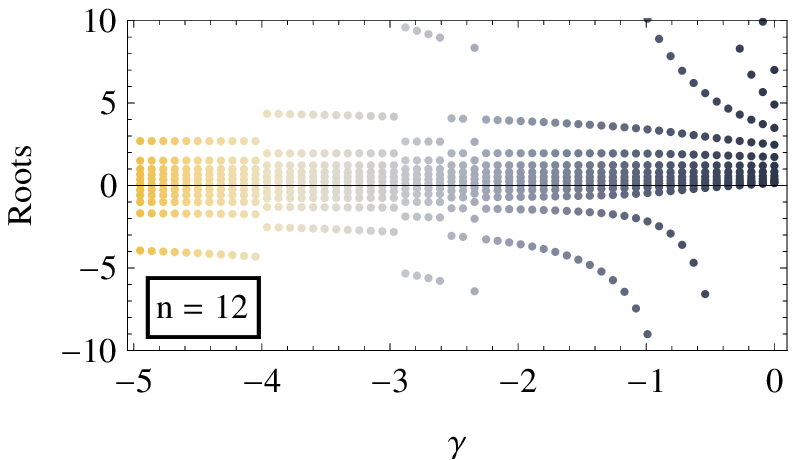}}
\par\end{centering}

\centering{}\caption{\label{fig:BAEsolutionN}Solutions of BAE (\ref{BAE})
for the ground state for $n=6$, $8$, $10$ and $12$, respectively.
We set the parameters $k$ and $\alpha$ to satisfy the condition
$\lambda=2,\,\beta=0$ in each case. There are abrupt changes in the
distribution of roots for all values of $n$. The changes occurs
at (a) $\gamma\simeq-2.30$ and $\gamma\simeq-2.75$; (b) $\gamma\simeq-2.25$,~$\gamma\simeq-2.54$
and $\gamma\simeq-3.12$, (c) $\gamma\simeq-2.25$, $\gamma\simeq-2.45$,
$\gamma\simeq-2.75$ and $\gamma\simeq-3.54$; (d) $\gamma\simeq-2.25$,
$\gamma\simeq-2.35$, $\gamma\simeq-2.54$, $\gamma\simeq-2.92$ and
$\gamma\simeq-3.98$. }
\end{figure*}

\par\end{center}
and by considering the terms of order $n$ in (\ref{eigenvalue equation}),
the energy eigenvalues are found to be 
\begin{equation}
E=\frac{kn^{2}}{4}+\frac{\alpha n}{2}-\frac{J}{2}\sum_{j=1}^{n}\upsilon_{j}-\Omega\sum_{j=1}^{n-1}\sum_{l=j+1}^{n}\upsilon_{j}\upsilon_{l}\label{eigenvalues}
\end{equation}
Each set of roots $\left\{ \upsilon_{j},\,j=1,\,...,\,n\right\} $
of the BAE leads to an energy of the Hamiltonian through  (\ref{eigenvalues}). 
Note that the change $J\rightarrow-J$ is equivalent
to the change $\upsilon_{j}\rightarrow\upsilon_{j}^{-1}$. For $\alpha=0$
this shows that each solution set $\left\{ \upsilon_{1},\,...,\,\upsilon_{n}\right\} $
is invariant under $\upsilon_{j}\rightarrow\upsilon_{j}^{-1}$. In
principle, an analytic solution of these equations is not possible.
Below, we implement numerical techniques to obtain solutions. 

We restrict ourselves to study the case $k>0$, $\alpha=0$ (due to
the relations (\ref{parameter correspondence}) this is equivalent
to $\lambda>0$, $\beta=0$) to investigate the behaviour of the BAE
solutions around the QPT line $\gamma=-1/2$. We start solving the
Bethe ansatz equations with $\Omega=0$ for the ground state. In this
case, all the roots must be real and positive \cite{LINKS2}. If we
decrease the value of $\Omega$, the numerical solution of the equations
(\ref{BAE}) shows that the ground state has always real roots, but
eventually some roots have a smooth transition from positive to negative
values. As some roots approach to zero, other ones diverge due the
invariance $\upsilon_{j}\rightarrow\upsilon_{j}^{-1}$. It must be
noted that this transition from positive to negative roots has no
relation with the QPT of this model.

In Figures \ref{fig:BAEsolutionN4} and \ref{fig:BAEsolutionN} we plot solutions of the BAE for certain values of the total number of particles $n$. These numerical solutions agree with 
the exact diagonalization of the Hamiltonian. Starting with Figure \ref{fig:BAEsolutionN4}, we plot
the solutions to the BAE (\ref{BAE}) with $n=4$. The roots generally evolve smoothly as the value
of the parameter $\gamma=n\Omega/J$ varies, although for some particular
values the trajectories exhibit jumps. This same
characteristic behavior of the ground state roots is observed for other values of $n$ - see Figure \ref{fig:BAEsolutionN}. 

\begin{center}
\begin{figure*}[t]
\begin{centering}
\subfloat[]{\begin{centering}
\includegraphics{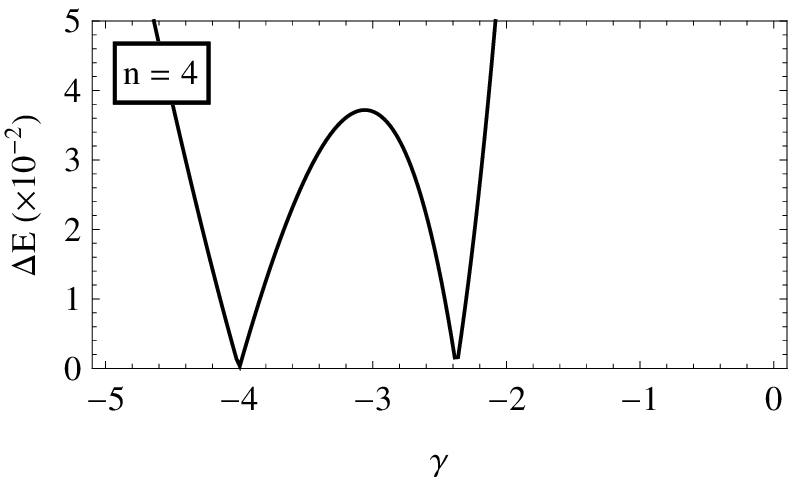}
\par\end{centering}

}~~~~~\subfloat[]{\centering{}\includegraphics{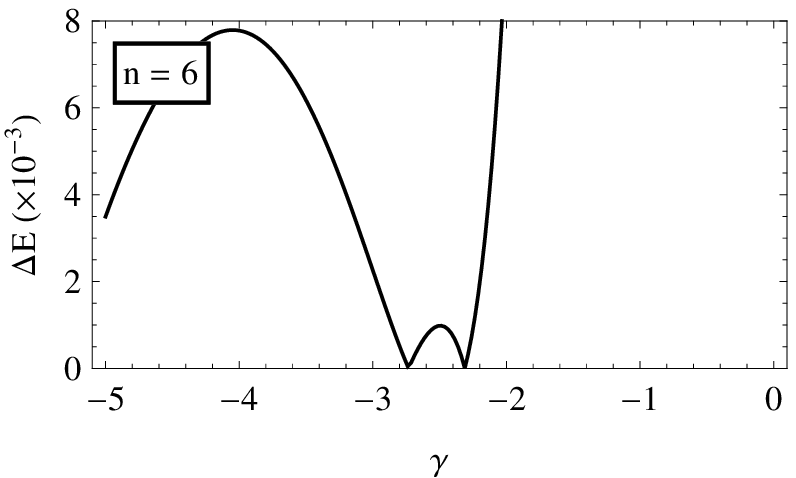}}\\
\subfloat[]{\centering{}\includegraphics{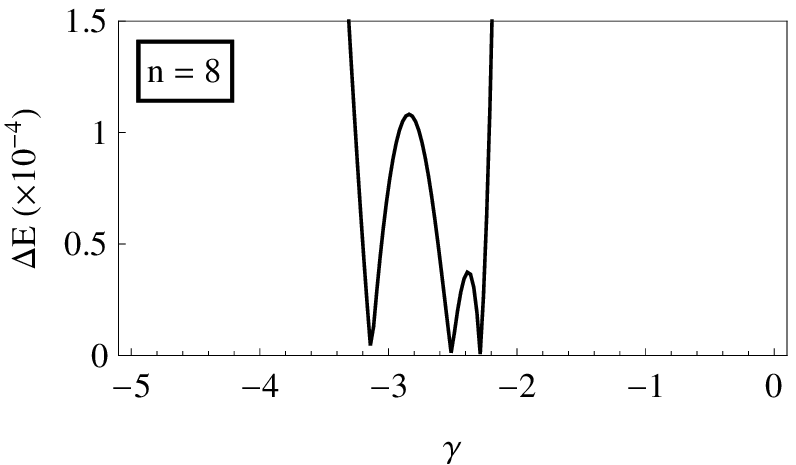}}~~~~~\subfloat[]{\centering{}\includegraphics{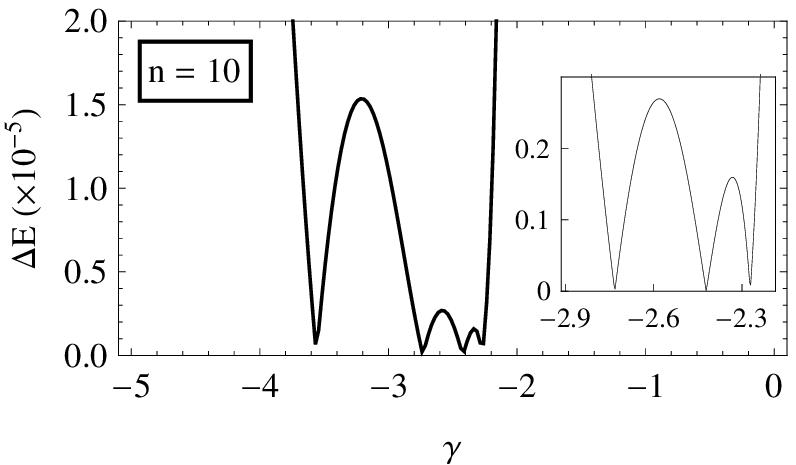}}\\
\subfloat[]{\centering{}\includegraphics{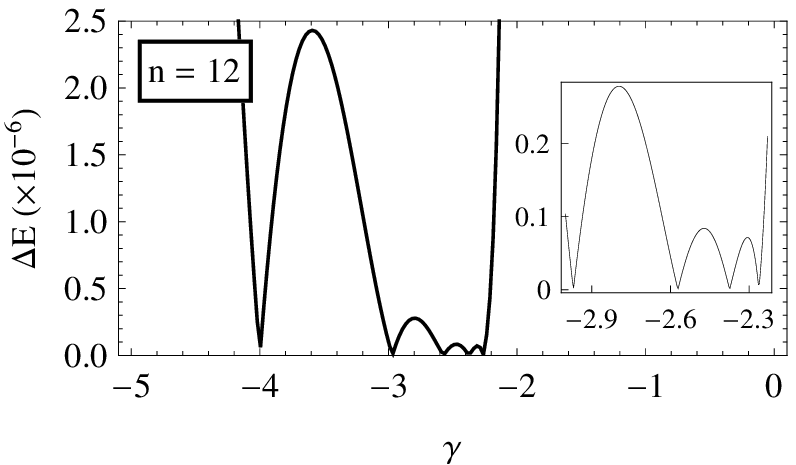}}
\par\end{centering}
\centering{}\caption{\label{fig:BAEsolutionN-1}Energy gap between the ground state and
the first excited state particular cases (a) $n=4$, (b) $n=6$, (c)
$n=8$, (d) $n=10$ and (e) $n=12$. We set the parameters $\lambda=2,\,\beta=0$
for all cases. The presence of non-zero regions in the energy gap
indicates that there are level crossing between the ground state and
the first excited state at some particular values of $\gamma$. }
\end{figure*}

\par\end{center}

Examination of the energy levels of the system for small number
of particles shows that there are crossings of levels between the
ground state and the first excited state, detected due to the presence
of non-zero regions in the energy gap. Note that the number of non-zero
regions in the energy gap increases along with the number of particles,
while it's amplitude becomes smaller (in fact, Figure \ref{fig:BAEsolutionN-1}
shows that the amplitute of the non-zero regions of the energy gap
becomes $10\times$ smaller every time we add two particles to the
system). 
%
%Also, from Figure \ref{fig:BAEsolutionN-1}, we identify
%the critical points of QPT for a given set of parameters and we note
%the striking correspondence between the points obtained by investigating
%the energy gap, that indicate a QPT, and the point where the respective
%ground state solutions of the BAE change their behavior. 
%
We also note
that, as the number of particles increases, the solutions of Bethe
ansatz equations still predict the crossing of energy levels, despite
the small value of $\Delta E$. 

%Therefore the behavior of the ground
%state of the model translated as the behavior of the set of solutions
%of the BAE, when the relevant parameters are changed, can be used
%as tool to understand the phase transition of the system. Similar
%correspondences have also been witnessed in other models admitting
%exact Bethe ansatz solutions. For example, by numerically solving
%the Bethe ansatz equations for the ground state, it was found in \cite{RUBENI}
%that there is a sharp change in the character of the root distribution
%in the complex plane around a particular coupling value. Through complementary
%computations of entanglement, fidelity, and the energy gap, it was
%identified that the change in the root distribution coincides with
%a quantum phase transition. By calculation of the ground-state Bethe
%root density in the limit of infinite number of particles the work
%\cite{LINKS3} obtained analytic expressions for the ground state
%energy which showed excellent agreement with numerical calculations.

\section{Summary}

In this work we introduced an eTMBH model with 
non-linear tunneling interaction term. We found that the model exhibits
QPT between three different phases: a Josephson phase, a self-trapping
phase and a phase-locking phase. This result was obtained through a classical analysis, 
allowing for the identification 
the parameter space of phase transitions as depicted
in Fig. \ref{fig:ParameteSpaceQPT}. For the case $\lambda>0$,
we compared the predictions coming from the classical analysis with the energy gap. It was found that the boundary between the 
Josephson and phase-locking regimes coincides with the closing of the gap.

%The main purpose of this study was to investigate the extended two-mode
%Bose-Hubbard model, an exactly solvable model in the Bethe ansatz
%sense, and explore how the presence of significant physical phenomena
%can be inferred from the Bethe ansatz structure. We began with a classical
%analysis for the model exploiting the fixed point structure and making
%explicit the presence of bifurcation points for critical values of
%the relevant parameters. 
%
%We have found that the eTMBH model may experience
%QPT between three different phases: a Josephson phase, a self-trapping
%phase and a phase-locking phase. Our analysis allowed us to build
%the parameter space of phase transitions for this model, as depicted
%in Fig. \ref{fig:ParameteSpaceQPT}. Considering the case $\lambda>0$,
%we compared the predictions coming from the parameter space of phase
%transitions with the energy gap and the trend unveiled by this method
%is highly compatible with the ground state results coming from the
%classical analysis. 
%
We then presented the exact solution for this model using the Bethe
ansatz method. 
%
%The BAEs thus obtained are quite involved and, apart
%from some limiting situations, it is unfeasible to obtain an analytical
%solution. Nevertheless the structure of the BAEs for the model allows
%the possibility of obtaining well behaved numerical solutions. The
%structure of these solutions present a peculiar behavior when some
%parameters of the Hamiltonian are varied, indicating that the ground
%state solutions of the BAEs are reflecting some change in the system
%energy spectrum behavior. 
%
Guided by the location of quantum phase transition boundaries predicted by
the classical analysis, we analysed solution of the
BAEs and the energy gap. Crossing of levels
between the ground state and the first excited state for a relatively
small number of particles were detected. As we increase the number of particles,
the crossings between these two levels becomes more frequent and with
smaller amplitude of $\Delta E$. The behaviour of the solutions for
the BAE change at the points where the energy gap goes to zero. 

The unusual features uncovered in this study call for a deeper analysis of the model. 
In future work it is planned to 
extend the methods adopted in \cite{RUBENI,LINKS3} for the TMBH model to meet this need.  
%
%Therefore
%the behavior of the ground state of the model translated as the behavior
%of the set of solutions of the BAE, when the relevant parameters are
%changed, can be helpful to locate the points of the phase transition
%of the system. We foresee the possibility of applying this kind of
%analysis in many different integrable models and this could possibly
%lead to some grouping according to the geometrical patterns formed
%by the roots such as arcs and lines, or eventually closed curves in
%other situations. 

\section{Acknowledgements}

Diefferson Rubeni and Angela Foerster are supported by CNPq (Conselho
Nacional de Desenvolvimento Científico e Tecnológico), Brazil. Jon
Links, Phillip Isaac and Angela Foerster are supported by the Australian Research Council
through Discovery Project DP150101294.

\end{document}